\documentclass[10pt,twocolumn,letterpaper]{article}
\usepackage[table,xcdraw]{xcolor}
\usepackage{caption}
\usepackage[utf8]{inputenc}
\usepackage{algorithm}
\usepackage{algpseudocode}
\usepackage{amsmath} 
\usepackage{booktabs}
\usepackage{multirow}
\usepackage{subcaption}
\usepackage{graphicx}
\usepackage{xcolor}
\usepackage[a4paper, margin=1in]{geometry}

\usepackage{booktabs}
\usepackage{siunitx}
\usepackage{cvpr}              

\definecolor{colorBlue}{rgb}{0.7, 0.85, 0.95}

\definecolor{colorRed}{rgb}{0.95, 0.7, 0.7}

\definecolor{colorGreen}{rgb}{0.7, 0.9, 0.8}

\definecolor{colorGray}{rgb}{0.85, 0.85, 0.85}

\definecolor{cvprblue}{rgb}{0.21,0.49,0.74}
\usepackage[pagebackref,breaklinks,colorlinks,citecolor=cvprblue]{hyperref}

\def\paperID{*****} 
\def\confName{ICCV}
\def\confYear{2025}

\title{Taming Domain Shift in Multi-source CT-Scan Classification\\ via Input-Space Standardization
}
\author{$^\ddagger$$^1$Chia-Ming Lee, $^1$Bo-Cheng Qiu, $^2$Ting-Yao Chen, $^1$Ming-Han Sun, $^2$Fang-Ying Lin \\$^1$Jung-Tse Tsai, $^2$I-An Tsai, $^1$Yu-Fan Lin,$^\dagger$$^{1,2}$Chih-Chung Hsu
\\
$^1$National Cheng Kung University \quad
$^2$National Yang Ming Chiao Tung University\\
{\tt\small $^\ddagger$zuw408421476@gmail.com, $^\dagger$cchsu@gs.ncku.edu.tw }
}

\begin{document}
\maketitle
 
\begin{abstract}
Multi-source CT-scan classification suffers from domain shifts that impair cross-source generalization. While preprocessing pipelines combining Spatial-Slice Feature Learning (SSFL++) and Kernel-Density-based Slice Sampling (KDS) have shown empirical success, the mechanisms underlying their domain robustness remain underexplored. This study analyzes how this input-space standardization manages the trade-off between local discriminability and cross-source generalization. The SSFL++ and KDS pipeline performs spatial and temporal standardization to reduce inter-source variance, effectively mapping disparate inputs into a consistent target space. This preemptive alignment mitigates domain shift and simplifies the learning task for network optimization. Experimental validation demonstrates consistent improvements across architectures, proving the benefits stem from the preprocessing itself. The approach's effectiveness was validated by securing first place in a competitive challenge, supporting input-space standardization as a robust and practical solution for multi-institutional medical imaging.
\end{abstract}
\vspace{-0.3cm}
\section{Introduction}

AI-assisted COVID-19 diagnosis through chest CT analysis faces a critical challenge: models trained on single-institution datasets often fail when deployed across different hospitals. While chest CT remains the gold standard for detecting COVID-19 manifestations~\cite{kollias2021mia,kollias2022ai,kollias2020deep,kollias2024sam2clip2sam}, systematic domain shifts across medical institutions fundamentally undermine cross-institutional generalization \cite{kollias2020transparent,kollias2024domain,kollias2023deep}.

\begin{figure}[ht!]
\centering
\includegraphics[width=1\linewidth]{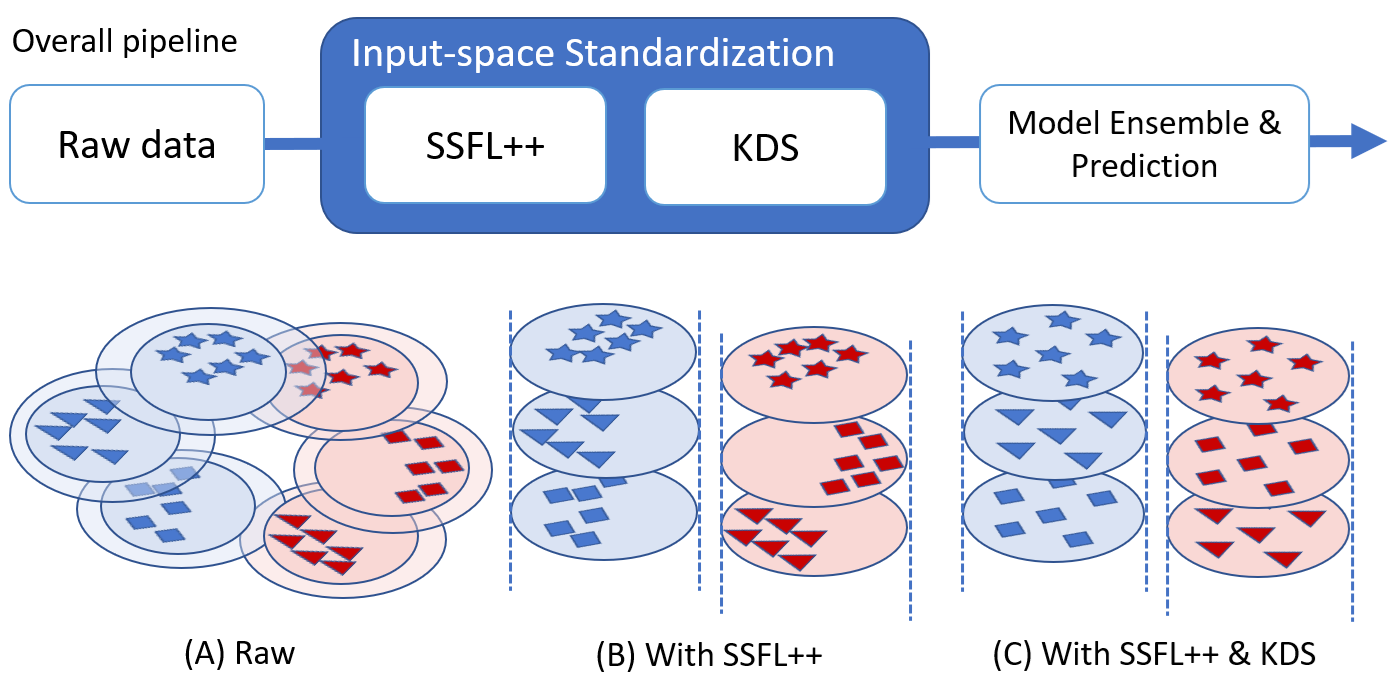}
\caption{Overview of the input-space standardization pipeline \cite{hsu2024closer,hsu2023bag}. SSFL++ ensures spatial alignment through lung-centric cropping, while KDS provides temporally consistent slice sampling based on anatomical density distributions. This study aims to analyze and demonstrate how standardized inputs facilitate strong cross-source generalization.
}
\label{fig:overall}
\end{figure}

These domain shifts stem from variations in reconstruction kernels, slice thickness, radiation dose optimization, and patient positioning protocols~\cite{guan2021domain,choudhary2020advancing}, creating a paradoxical situation where COVID-positive scans from different hospitals exhibit greater dissimilarity than COVID-negative scans from the same institution.

Contemporary approaches predominantly rely on post-hoc domain adaptation techniques that attempt to disentangle domain artifacts after feature learning. These include methods like domain adversarial training~\cite{ganin2016domain}, invariant risk minimization (IRM)~\cite{arjovsky2019invariant, krueger2021outofdistributiongeneralizationriskextrapolation}, and its variants like Variance Risk Extrapolation (VREx)~\cite{yuan1}. However, these reactive methods face significant limitations in resource-constrained medical environments where domain labels are unavailable and computational budgets impose practical constraints~\cite{li2024enhancing, wang2021tent, guan2022domainatmdomainadaptationtoolbox, laparra2020rethinking}.

To overcome these limitations, our work shifts the focus from post-hoc correction to preemptive prevention. We propose an alternative paradigm through input-space standardization, adopting the preprocessing pipeline of Spatial-Slice Feature Learning (SSFL++) and Kernel-Density-based Slice Sampling (KDS) introduced by Hsu et al.~\cite{hsu2024closer, hsu2023bag}. This preventive approach performs spatial and temporal standardization to reduce inter-source variance, effectively mapping disparate inputs into a consistent target space. By aligning the data before feature learning, this process is designed to mitigate domain shift and simplify the network optimization. Our primary contribution is a comprehensive feature space analysis that characterizes how this standardization pipeline manages the fundamental trade-off between local discriminability and cross-domain generalization, a mechanism that was previously underexplored.
In summary, our contribution can be listed as follows:
\begin{itemize}
    \item \textbf{Analysis of a Standardization Trade-off}: We provide a quantitative analysis of the trade-off between local class separation and cross-domain consistency. Our analysis shows that the complete pipeline mitigates the domain-specific overfitting seen with spatial standardization alone, reducing inter-source variance by 75\%.

    \item \textbf{Architecture-Independent Validation}: We demonstrate that the performance improvements are consistent across distinct model architectures (a CNN and a Transformer). This supports the conclusion that the gains are primarily attributable to the data-centric preprocessing, not model-specific biases.

    \item \textbf{A Practical and Competitive Methodology}: We present a computationally efficient preprocessing pipeline suitable for clinical deployment. Its competitiveness was demonstrated by achieving first place in the PHAROS-AFE-AIMI Competition.
\end{itemize}

The remainder of this paper presents related work (Section \ref{related}), our SSFL++ and KDS methodology (Section \ref{sec:method}), comprehensive analysis for methods (Section \ref{sec:analysis}), experimental results including competition performance (Section \ref{sec:experiments}), and conclusions (Section \ref{sec:conclusion}).
\section{Related Work}
\label{related}
\subsection{Domain Adaptation in Medical Imaging}

Medical imaging faces persistent domain shift challenges from varied acquisition protocols, scanner manufacturers, and institutional practices~\cite{choudhary2020advancing, guan2021domain,lin2024divideconquergroundingbleeding}, which can cause performance to drop by 20-40\% when models are deployed across different hospitals~\cite{ayana2024multistage, abid2023multi}.

The dominant paradigm to address this is \textbf{post-hoc domain adaptation}, where methods operate in the feature space \textit{after} learning. Techniques like domain adversarial training~\cite{ganin2016domain} or IRM~\cite{arjovsky2019invariant, krueger2021outofdistributiongeneralizationriskextrapolation} attempt to learn domain-invariant features. However, these reactive approaches are often impractical for clinical deployment because they can be computationally expensive and typically require domain labels, which are often unavailable~\cite{laparra2020rethinking, li2024enhancing, guan2022domainatmdomainadaptationtoolbox}.

An alternative, preventive paradigm is \textbf{input-space standardization}, which harmonizes data \textit{before} feature learning~\cite{abbasi2024deep, karani2018lifelonglearningapproachbrain}. This field includes post-processing frameworks, often based on Generative Adversarial Networks (GANs) like STAN-CT~\cite{selim2020stanctstandardizingctimage} and RadiomicGAN~\cite{9669448}, that standardize images from single (intra-scanner) or multiple (cross-scanner) devices~\cite{selim2021crossvendorctimagedata}. While these GAN-based methods have laid a foundation, they can have limitations, such as being restricted to image patch synthesis~\cite{8904763}. In contrast, the direct preprocessing pipelines central to our work, SSFL++ and KDS~\cite{hsu2024closer, hsu2023bag}, have shown strong empirical success. However, the mechanisms that enable their domain robustness remain analytically underexplored, motivating this study.

\subsection{COVID-19 CT Analysis and Detection}

In recent years, significant advancements have been made in devising techniques for COVID-19 identification. Kollias \emph{et al}. \cite{kollias2020deep,kollias2021mia,kollias2022ai} have advanced this area by evaluating the outcomes of deep learning models that rely on latent representations. Chen \emph{et al}. \cite{chen2021adaptive} employed maximum likelihood estimation combined with the Wilcoxon test, using a statistical learning approach to adaptively choose slices and create explainable models.

Moreover, Hou \emph{et al}. developed a method utilizing contrastive learning to improve feature representation \cite{hou1,hou2}. Turnbull \emph{et al}. implemented a 3D ResNet \cite{ResNet} for classifying COVID-19 severity. Hsu \emph{et al}. \cite{hsu2022spatiotemporal} presented a strategy that integrates 2D feature extraction with an LSTM \cite{hochreiter1997long} and Vision Transformer \cite{dosovitskiy2020image} in a two-step model \cite{hsu2023bag}, achieving outstanding performance. Yuan \emph{et al}. incorporated VREx into the training process \cite{yuan1}, encouraging the model to maintain performance across multiple source domains by explicitly minimizing the variance.

\section{Methods}
\label{sec:method}

Our methodology tackles domain shift using a preventive, input-space standardization framework. Multi-source CT classification often suffers from systematic inconsistencies that go beyond acquisition protocols, including variations in spatial framing and slice selection strategies~\cite{hsu2024closer, hsu2023bag}. These variations introduce domain-specific artifacts and variance that undermine a model's ability to generalize across different hospitals.

To address these challenges at their source, our preprocessing framework is comprised of two complementary modules: \textbf{SSFL++} for spatial standardization and \textbf{KDS} for temporal standardization. By standardizing input characteristics before feature learning begins, our approach preemptively removes domain-related noise, rather than attempting to correct for it post-hoc.

\subsection{Input-Space Standardization Framework}

Traditional domain adaptation approaches operate in feature space, attempting to align learned representations after feature extraction~\cite{ganin2016domain,arjovsky2019invariant,drct,densesr}. This reactive paradigm suffers from a fundamental limitation: it addresses domain shifts after models have been exposed to heterogeneous inputs. In resource-constrained medical environments where annotated data is scarce, such post-hoc approaches often prove impractical~\cite{li2024enhancing,wang2024tent,mm24,prompthsi}.

Our framework is built on a preventive paradigm that standardizes input data before any feature learning occurs. Formally, given a set of input distributions $\{\mathcal{X}_1, \mathcal{X}_2, ..., \mathcal{X}_k\}$ from multiple sources, our objective is to find a transformation that maps them into a standardized representation space. This transformation is designed to preserve salient diagnostic information while removing source-specific artifacts, as illustrated by the pipeline architecture in Figure~\ref{fig:overall}.

\subsection{SSFL++: Spatial Standardization}

\begin{figure}[h]
\centering
\includegraphics[width=0.95\linewidth]{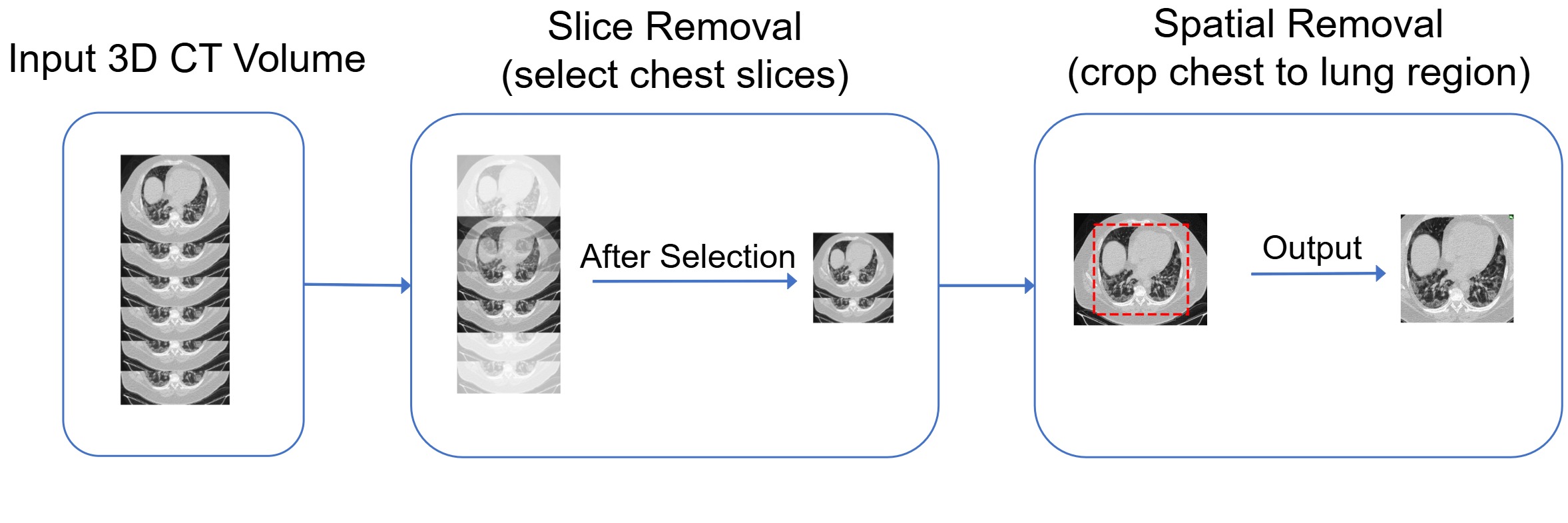}
\caption{Spatial standardization via SSFL++. \textbf{Left}: Original CT slice with source-specific field-of-view and background content. \textbf{Right}: Standardized lung-centric region with consistent anatomical focus, eliminating spatial domain shifts while preserving diagnostic information.}
\label{fig:cropping}
\end{figure}

SSFL++ addresses spatial domain shifts by enforcing consistent anatomical framing across all CT scans, regardless of their original field-of-view or scanning protocols. The spatial standardization process operates through systematic lung region extraction and normalization. Figure~\ref{fig:cropping} demonstrates the spatial standardization process.

The algorithm applies morphological filtering for noise reduction, followed by adaptive binarization to isolate lung regions:
\begin{equation}
\text{Mask}[i,j] = \begin{cases}
1, & \text{if } Z_{\text{filtered}}[i,j] \geq t \\
0, & \text{otherwise}
\end{cases}
\end{equation}
where threshold $t$ is determined adaptively for each scan. The union of lung masks across all slices defines the minimal bounding box encompassing lung anatomy, ensuring consistent anatomical focus while eliminating source-specific background variations. 

\subsection{KDS: Temporal Standardization}

KDS addresses temporal heterogeneity arising from variable scan lengths and slice selection protocols across sources. Rather than employing ad-hoc sampling strategies, KDS uses principled density-based selection to ensure consistent anatomical coverage.

Given lung tissue area measurements across slices, we estimate the anatomical density distribution using kernel density estimation. The cumulative distribution function enables percentile-based sampling:
\begin{equation}
F(q_p) = p, \quad p \in \{0.05, 0.15, ..., 0.85, 0.95\}
\end{equation}

This approach selects 8 slices at fixed percentiles of the anatomical density distribution, ensuring representative coverage of lung anatomy from apex to base, regardless of original scan characteristics.

\begin{figure}[h]
\centering
\includegraphics[width=0.99\linewidth]{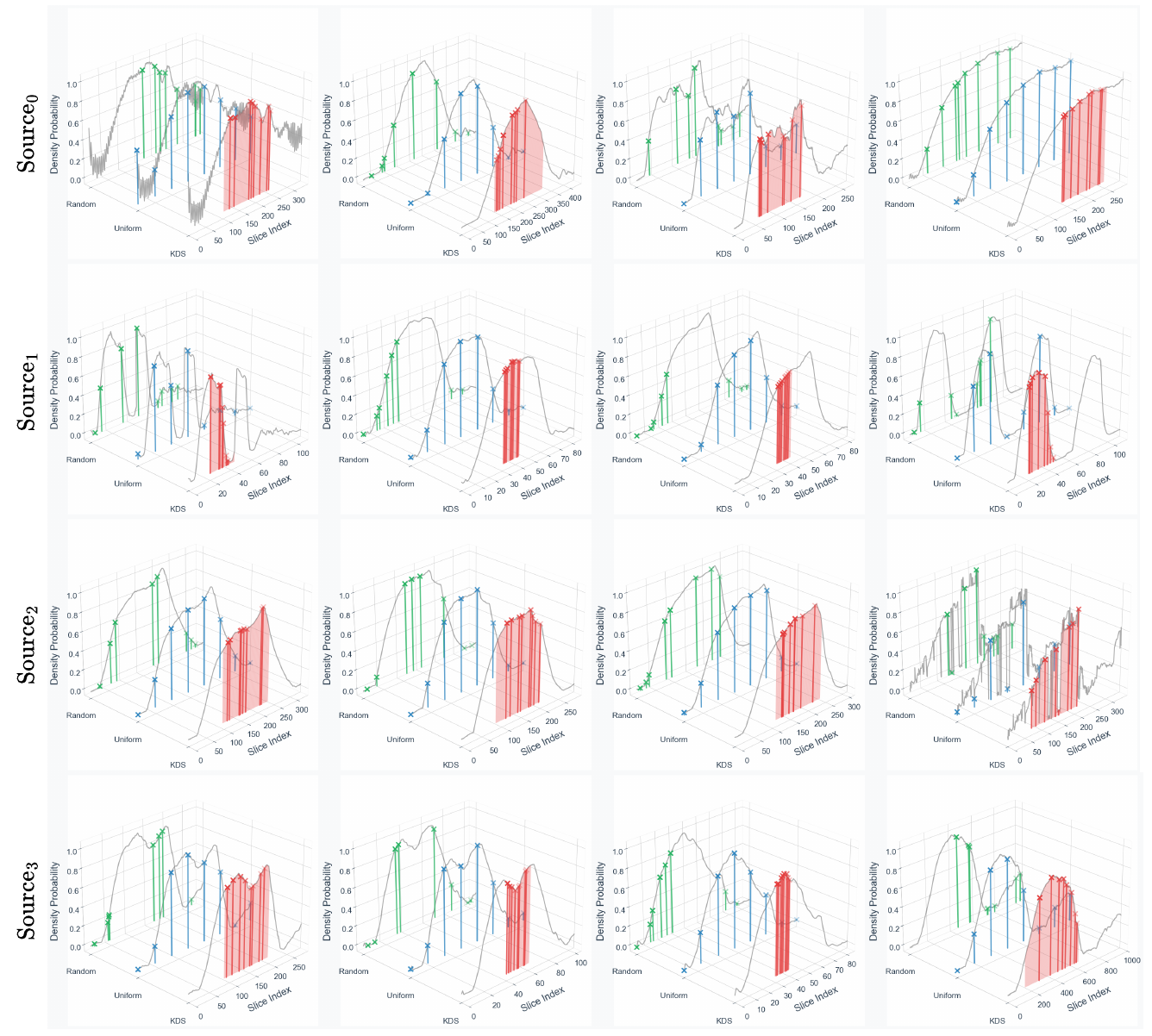}
\caption{Synergistic SSFL++ and KDS preprocessing visualization across four institutional sources. Lung tissue density distributions (gray curves) vary between institutions and scans, with slice selection methods shown as: \colorbox{colorRed}{KDS}, \colorbox{colorBlue}{uniform}, and \colorbox{colorGreen}{random} sampling. The red planes covering KDS sampling points represent SSFL++ spatial boundaries, demonstrating how spatial standardization enables more representative temporal sampling.}
\label{fig:sampling}
\end{figure}

The synergistic effect of SSFL++ and KDS is illustrated in Figure~\ref{fig:sampling}. Lung tissue density distributions vary significantly between institutions, highlighting temporal inconsistency challenges. While random and uniform sampling fail to account for anatomical variations, KDS adapts to each scan's density profile, ensuring consistent coverage. The spatial boundaries established by SSFL++ provide a consistent anatomical reference frame that enables more effective temporal sampling across all sources.

\subsection{Classification Architecture}

The standardized input volumes are processed through an EfficientNet-B3~\cite{efficientnet} that extracts features from the 8 standardized slices. The network architecture employs slice-level feature averaging followed by binary classification:

\begin{align}
\mathbf{f}_{\text{agg}} &= \frac{1}{8} \sum_{i=1}^{8} \text{EfficientNet-B3}(\mathbf{s}_i) \\
p &= \text{Sigmoid}(\text{Linear}(\mathbf{f}_{\text{agg}}))
\end{align}

where $\mathbf{s}_i$ represents the $i$-th standardized slice and $p \in [0,1]$ is the final COVID-19 prediction probability. This approach leverages information from all selected slices while maintaining computational efficiency, distinguishing between COVID-19 positive and negative cases based on the aggregated representation.

The following analysis investigates how these preprocessing components affect feature representations and cross-source generalization.

\section{Analysis}
\label{sec:analysis}

Having established our preprocessing methodology, this section presents a quantitative analysis of its impact on the learned feature representations. The goal is to dissect the mechanism by which our input-space standardization manages the fundamental trade-off between \textbf{local discriminative capability} (a model's performance within a single source) and \textbf{cross-domain generalization} (its robustness across different sources).

\subsection{Analytical Framework}

Our analytical framework is designed to evaluate whether pre-emptively standardizing inputs is more effective than correcting for domain shift post-hoc. To do this, we examine feature embeddings, $f(x) \in \mathbb{R}^{d}$, extracted from the penultimate layer of a trained EfficientNet-B3 classifier~\cite{efficientnet}.

We then apply a set of specialized metrics (detailed in Section~\ref{sec:eval_metrics}) to these embeddings. These metrics are designed to explicitly quantify the two competing objectives:
\begin{itemize}
    \item \textbf{Discriminative Capability}, which measures how effectively the features separate classes within each domain.
    \item \textbf{Cross-Source Consistency}, which measures the alignment of feature distributions for the same class across different domains.
\end{itemize}
This framework allows us to systematically characterize how each component of our preprocessing pipeline influences the balance between these two objectives and ultimately contributes to robust, generalizable performance.

\begin{table*}[h]
\centering
\caption{Analysis across preprocessing strategies. Higher Fisher Score and Separability indicate better discriminative capability. Lower Domain Consistency and Inter-source Variance indicate better cross-source alignment.}
\label{tab:feature_analysis}
\small
\begin{tabular}{lccccc}
\toprule
\multirow{3}{*}{\textbf{Method}} & \multicolumn{2}{c}{\textbf{Discriminative Capability} $\uparrow$} & \multicolumn{3}{c}{\textbf{Cross-Source Consistency} $\downarrow$} \\
\cmidrule(lr){2-3} \cmidrule(lr){4-6}
& \textbf{Fisher Score} & \textbf{Separability} & \textbf{Domain Consistency} & \textbf{Inter-source Variance} & \textbf{Inter-source Variance} \\
& & & & \textbf{(COVID)} & \textbf{(Non-COVID)} \\
\midrule
w/ KDS & 2.14 & 1.45 & 0.20 & 0.87 & 0.51 \\
w/o KDS & 2.06 & 3.55 & 1.91 & 3.49 & 1.76 \\
Baseline & 1.81 & 1.44 & 0.22 & 1.06 & 1.33 \\
\bottomrule
\end{tabular}
\end{table*}
\subsection{Evaluation Metrics}
\label{sec:eval_metrics}

To quantitatively measure the fundamental trade-off between a model's discriminative power and its ability to generalize, we define a set of complementary metrics. These metrics are designed to evaluate the feature space organization from two key perspectives: class separability and cross-source consistency.

\subsubsection{Discriminative Capability Metrics}
These metrics assess how well the learned features can separate the different classes.

\begin{itemize}
    \item \textbf{Fisher Score}: Measures the overall, global separability between the COVID and non-COVID classes across all sources. A higher score indicates better class separation.
    \begin{equation}
        \text{Fisher Score} = \frac{\|\mu_{\text{COVID}} - \mu_{\text{non-COVID}}\|_2}{\frac{1}{2}(\bar{d}_{\text{COVID}} + \bar{d}_{\text{non-COVID}})}
    \end{equation}

    \item \textbf{Separability}: Measures the average local, per-domain class separation. This allows us to assess if a model performs well within a specific source, even if it doesn't generalize.
    \begin{equation}
        \text{Separability} = \frac{1}{S} \sum_{s=1}^{S} \frac{\|\mu_{s,\text{COVID}} - \mu_{s,\text{non-COVID}}\|_2}{\bar{d}_{s,\text{COVID}} + \bar{d}_{s,\text{non-COVID}}}
    \end{equation}
\end{itemize}

\subsubsection{Cross-Source Consistency Metrics}
This metric assesses how well the feature representations for a given class align across different source domains, which is a proxy for generalization capability.

\begin{itemize}
    \item \textbf{Inter-Source Variance}: Quantifies the distribution shift for a single class ($c$) by measuring the average distance between its feature centroids from different source domains ($i$ and $j$). A lower value indicates that the feature representations are more consistent and generalizable.
    \begin{equation}
        \text{Inter-Source Variance}_c = \frac{1}{\binom{S}{2}} \sum_{i<j} \|\mu_{i,c} - \mu_{j,c}\|_2
    \end{equation}
\end{itemize}

In these equations, $S$ is the total number of sources, and $\mu_{s,c}$ is the feature centroid for class $c$ in source $s$. The term $\bar{d}_{s,c}$ represents the average intra-class pairwise distance, calculated as:
\begin{equation}
    \bar{d}_{s,c} = \frac{1}{\binom{N_{s,c}}{2}} \sum_{x_a, x_b \in \mathcal{D}_{s,c}} \|f(x_a) - f(x_b)\|_2
\end{equation}
where $\mathcal{D}_{s,c}$ is the set of samples for class $c$ from source $s$, $N_{s,c}$ is the number of such samples, and $f(x)$ is the feature embedding.

\begin{figure}[h]
\centering
\includegraphics[width=1.05\linewidth]{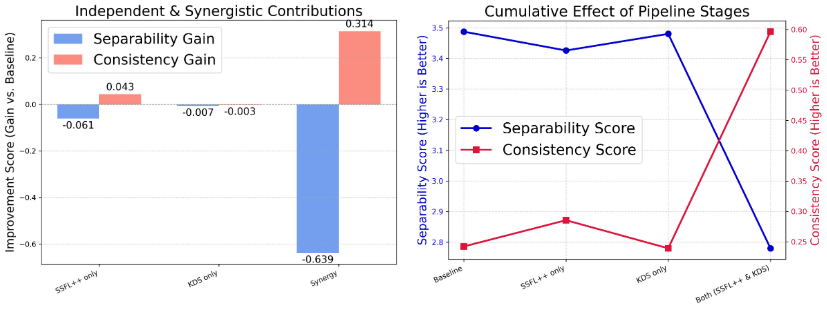}
\caption{Pipeline component analysis and cumulative effects. \textbf{Left}: Individual and synergistic contributions showing SSFL++ provides modest gains while KDS alone has minimal impact, but their combination yields substantial consistency improvements (-0.639). \textbf{Right}: Cumulative progression reveals the separability-consistency trade-off: separability decreases while consistency improves, with the complete pipeline achieving optimal cross-source balance.}
\label{fig:decomp}
\end{figure}

\subsection{Results and Interpretation}

Our feature space analysis reveals a fundamental trade-off in multi-source learning: models optimized for high discriminability within a single source often learn institutional artifacts, leading to poor cross-source consistency and generalization. Input-space standardization constrains this behavior by pre-emptively removing source-specific cues before feature learning occurs.

The analysis further demonstrates a strong synergistic effect between our pipeline's components. While spatial standardization (SSFL++) alone provides a modest improvement, our results show that a significant gain in cross-source consistency is only achieved when it is combined with temporal standardization (KDS). This suggests that establishing a consistent anatomical frame is a necessary foundation for effective, density-based slice selection to work reliably across domains.

\begin{figure}[h]
\centering
\includegraphics[width=0.99\linewidth]{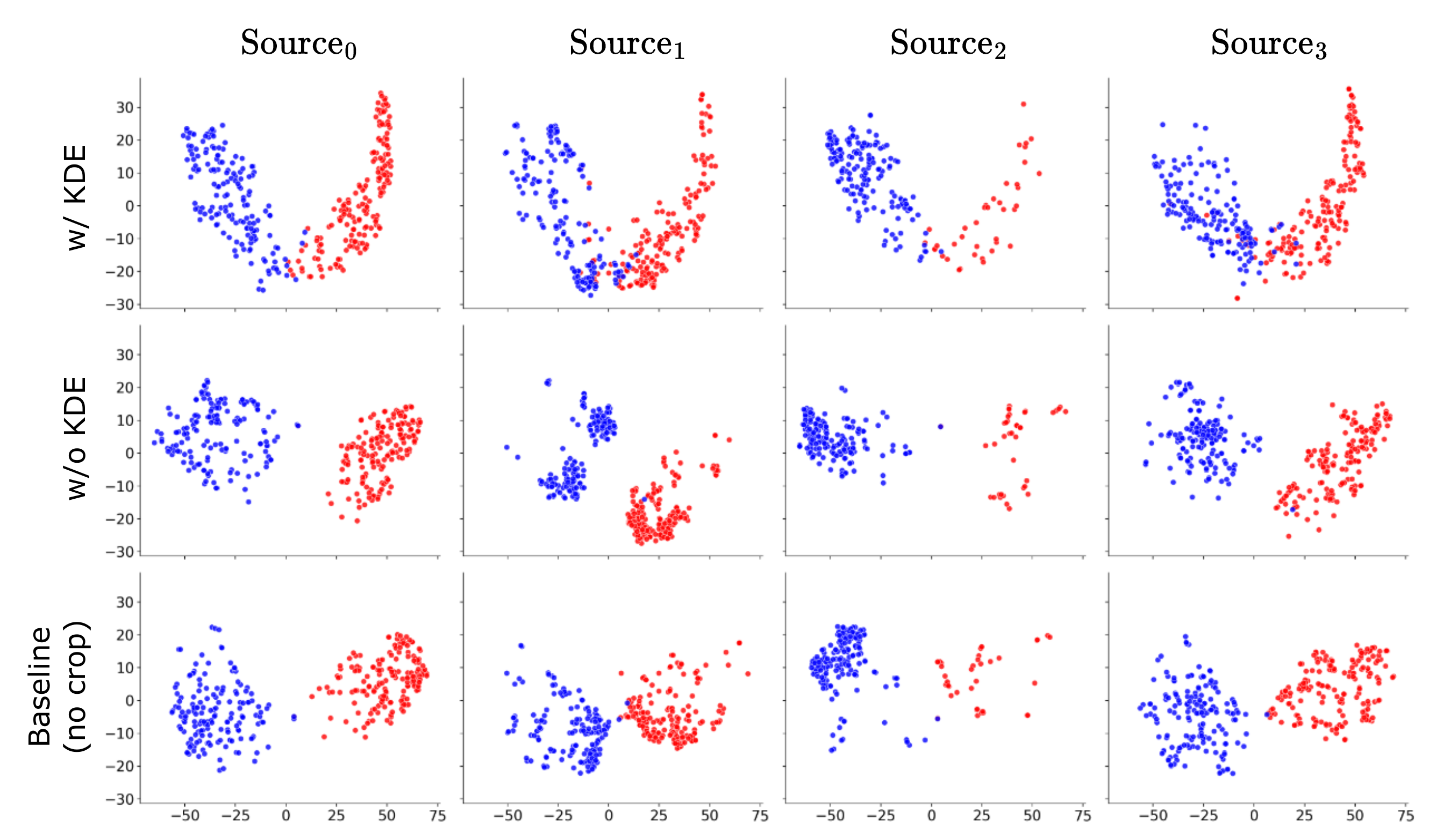}
\caption{Multi-source t-SNE visualization of learned feature representations across four sources. \textbf{Top}: Model with KDS preprocessing shows tight clusters with strong class separation and cross-domain consistency. \textbf{Middle}: Model without KDS shows dispersed representations with visible source-specific clustering. \textbf{Bottom}: Baseline model displays poor class separation and notable inter-source overlap. Colors represent classes (COVID/non-COVID) and shapes represent different sources.}
\label{fig:tsne}
\end{figure}

\begin{figure*}[ht!]
\centering
\includegraphics[width=1\linewidth]{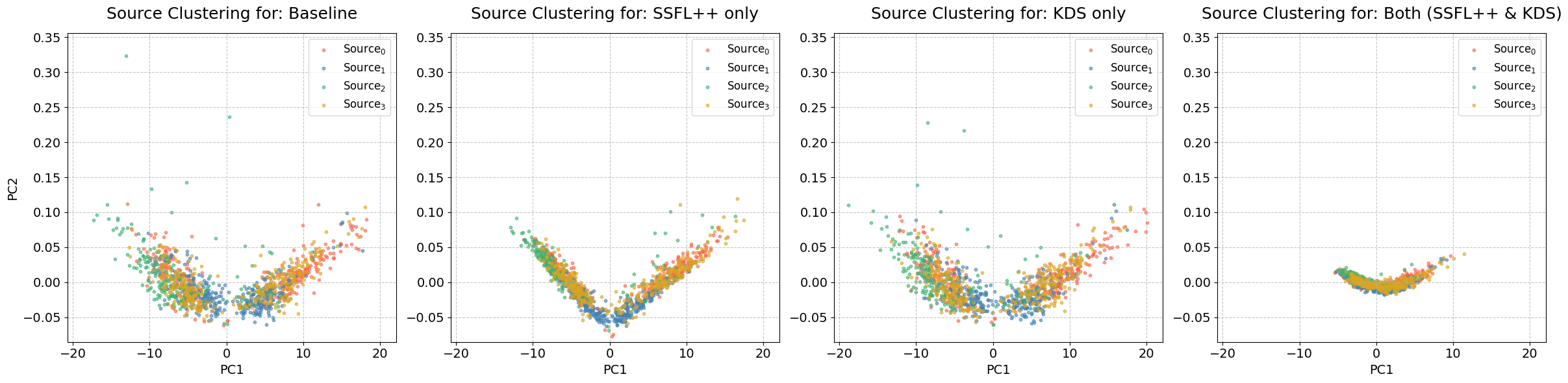}
\caption{Source clustering analysis through PCA visualization of learned feature representations across four  sources under different preprocessing configurations.  \textbf{Left}: Baseline model without preprocessing shows mixed clustering with overlapping source distributions. \textbf{Center-left}: SSFL++ only configuration demonstrates improved spatial consistency but maintains source-specific clustering patterns. \textbf{Center-right}: KDS only preprocessing achieves moderate source alignment with reduced inter-source scatter. \textbf{Right}: Combined SSFL++ and KDS preprocessing produces the most compact and source-agnostic feature representations, indicating effective input-space standardization.
}
\label{fig:pca}
\vspace{-0.3cm}
\end{figure*}

Visualizations of the feature space (Figures~\ref{fig:tsne} and~\ref{fig:pca}) confirm these findings. The combined pipeline successfully aligns feature clusters from different sources while maintaining clear separation between diagnostic categories. This result supports our core insight: that input standardization acts as an effective inductive bias. By enforcing spatial and temporal consistency at the data level, our pipeline constrains the model to learn features that are inherently more invariant to source-specific variations, a preventive approach that differs fundamentally from post-hoc correction methods.

These findings suggest that domain robustness in medical imaging may be more effectively achieved through careful input curation that addresses the root causes of domain shift—such as anatomical and temporal inconsistency—rather than relying on more complex algorithmic interventions after the fact.

\section{Experiments}
\label{sec:experiments}

This section experimentally evaluates our proposed preprocessing pipeline, SSFL++ and KDS, validating the feature space analysis presented in Section~\ref{sec:analysis}.

\subsection{Dataset}
We conduct experiments on the COVID-19-CT-DB dataset, a multi-source collection containing 2,899 CT scans from four different medical institutions. The distribution of scans across sources, which highlights the inherent domain adaptation challenges, is detailed in Table~\ref{tab:dataset_distribution}.

\begin{table}[h]
\centering
\caption{Dataset distribution across four medical sources.}
\label{tab:dataset_distribution}
\small
\begin{tabular}{lrrrr}
\toprule
\multirow{2}{*}{\textbf{Source}} & \multicolumn{2}{c}{\textbf{Scan Counts}} & \multicolumn{2}{c}{\textbf{Slice Counts}} \\
\cmidrule(lr){2-3} \cmidrule(lr){4-5}
& \textbf{COVID} & \textbf{Non-COVID} & \textbf{COVID} & \textbf{Non-COVID}  \\
\midrule
0 & 218 & 209 & 59,697 & 57,898 \\
1 & 218 & 210 & 10,213 & 13,802 \\
2 & 39 & 210 & 10,771 & 61,129  \\
3 & 217 & 210 & 38,573 & 35,916  \\
\midrule
Testing & \multicolumn{2}{c}{--} & \multicolumn{2}{c}{--} \\
\midrule
\textbf{Total} & \textbf{692} & \textbf{839} & \textbf{119,254} & \textbf{168,745}\\
\bottomrule
\end{tabular}
\end{table}

\subsection{Implementation Details}
Our primary classification architecture is an EfficientNet-B3~\cite{efficientnet}, trained using the Adam optimizer with a learning rate of 1e-4 and a weight decay of 5e-4. The model is trained to minimize the binary cross-entropy (BCE) loss, which is defined as:
\begin{equation}
    \mathcal{L}_{\text{BCE}} = -[y \log(p) + (1 - y) \log(1 - p)]
\end{equation}
where $y \in \{0, 1\}$ is the ground-truth label (non-COVID/COVID), and $p$ is the model's predicted probability for the positive class.

To improve model robustness, we also employ several data augmentation techniques during training, including random horizontal flipping, rotation (up to $\pm$30$^{\circ}$), and minor intensity variations.

\subsection{Evaluation Protocol}
\label{sec:eval_protocol}

We use a 5-fold cross-validation strategy for all experiments. The primary performance metrics are the macro-averaged F1-score and the Area Under the Receiver Operating Characteristic Curve (AUC-ROC).

\noindent\textbf{Macro F1-Score} is the unweighted mean of the F1-scores calculated for each class ($c \in C$):
\begin{equation}
    \text{Macro F1} = \frac{1}{|C|} \sum_{c \in C} \left( 2 \cdot \frac{\text{Precision}_c \cdot \text{Recall}_c}{\text{Precision}_c + \text{Recall}_c} \right)
\end{equation}

\noindent\textbf{AUC-ROC} measures the area under the curve formed by plotting the True Positive Rate (TPR) against the False Positive Rate (FPR):
\begin{align}
    \text{TPR} = \frac{\text{TP}}{\text{TP} + \text{FN}}, \quad \text{FPR} = \frac{\text{FP}}{\text{FP} + \text{TN}}
\end{align}

\subsection{Ablation Study}

\begin{table}[h]
\centering
\caption{Ablation study showing individual component contributions, evaluated with EfficientNet-B3 and Swin Transformer.}
\label{tab:component_analysis}
\footnotesize
\begin{tabular}{ccccc}
\toprule
\textbf{SSFL++} & \textbf{KDS} & \textbf{F1-score (\%)} & \textbf{AUC-ROC} & \textbf{Improvement} \\
\midrule
& & 70.73 & 0.7804 & -- \\
\checkmark & & 80.49 & 0.8957 & +9.76\% \\
\checkmark & \checkmark & 94.68 & 0.9813 & +23.95\% \\
\midrule
& & 68.53 & 0.7654 & -- \\
\checkmark & & 78.82 & 0.8672 & +10.82\% \\
\checkmark & \checkmark & 93.34 & 0.9797 & +24.81\% \\
\bottomrule
\end{tabular}
\vspace{-0.3cm}
\end{table}
Table~\ref{tab:component_analysis} evaluates the individual and combined contributions of the SSFL++ and KDS components across both the EfficientNet-B3 and Swin Transformer architectures.


The ablation results reveal a clear and consistent hierarchical contribution pattern across both architectures. For EfficientNet-B3 \cite{efficientnet}, spatial standardization (SSFL++) provides a foundational +9.76\% F1-score improvement, while the full pipeline yields a +23.95\% total gain. This pattern is mirrored on the Swin Transformer \cite{swin}, with SSFL++ contributing +10.82\% and the complete pipeline achieving a +24.81\% improvement.

The non-additive nature of these gains points to a strong synergy between the components. This performance hierarchy directly aligns with our analysis: SSFL++ alone improves results by enforcing spatial consistency, but our analysis shows this is insufficient as models still learn source-specific artifacts. The more substantial gain from KDS validates our finding that temporal heterogeneity is the more critical component of domain shift. KDS directly mitigates this primary source of error, which explains its larger contribution.

Ultimately, the study confirms that while both spatial and temporal standardization are required, addressing temporal consistency is the dominant factor for achieving robust, cross-source generalization. 

\subsection{Architecture Independence Validation}

In domain adaptation research, it's crucial to determine if improvements are due to preprocessing or architecture-specific optimizations. To clarify this, we assess our SSFL++ and KDS pipeline using various network architectures, including EfficientNet-B7 \cite{efficientnet}, Swin Transformer \cite{swin} and ResNet-101 \cite{ResNet}.

Table~\ref{tab:architecture_validation} demonstrates consistent improvements across different network architectures, validating that benefits stem from preprocessing rather than architecture-specific optimizations.

\begin{table}[h]
\centering
\caption{Performance consistency across different architectures.}
\label{tab:architecture_validation}
\small
\begin{tabular}{lcc}
\toprule
\textbf{Architecture} & \textbf{F1-score (\%)} & \textbf{AUC-ROC}\\
\midrule
EfficientNet-B3~\cite{efficientnet} & 94.68 & 0.9813  \\
Swin Transformer~\cite{swin} & 93.34 & 0.9797  \\
ResNet-101~\cite{ResNet} & 91.07 & 0.9421  \\
\bottomrule
\end{tabular}
\vspace{-0.1cm}
\end{table}

The narrow performance range across CNN and Transformer paradigms provides strong evidence for architecture independence. This consistency is particularly notable given the fundamental differences in feature learning mechanisms: EfficientNet-B3 employs convolutional feature extraction with compound scaling, while Swin Transformer utilizes self-attention mechanisms with hierarchical feature representations.

\subsection{Cross-Source Performance}

Table~\ref{tab:source_analysis} examines performance across individual sources and demonstrates practical effectiveness through competition results.

\begin{table}[h]
\centering
\caption{Cross-source performance and competition result.}
\label{tab:source_analysis}
\small
\begin{tabular}{cccc}
\toprule
\multirow{2}{*}{\textbf{Source}} & \multicolumn{2}{c}{\textbf{Validation Set}} & \multicolumn{1}{c}{\textbf{Competition Test}} \\
\cmidrule(lr){2-4} 
& \textbf{F1-score (\%)} & \textbf{AUC-ROC} & \textbf{F1-score (\%)} \\
\midrule
0 & 97.53 & 0.9981 & 94.2 \\
1 & 87.96 & 0.9564 & 74.2 \\
2 & 49.21$^*$ & --$^*$ & 51.2  \\
3 & 90.90 & 0.9672 & 90.9  \\
\midrule
\multicolumn{3}{c}{\textbf{Competition Result}} & \textbf{77.6\% (1st Place)} \\
\bottomrule
\end{tabular}
\vspace{0.1cm}
\small

$^*$ Results affected by validation set class imbalance.
\vspace{-0.1cm}
\end{table}

\begin{figure}[h]
\centering
\includegraphics[width=0.99\linewidth]{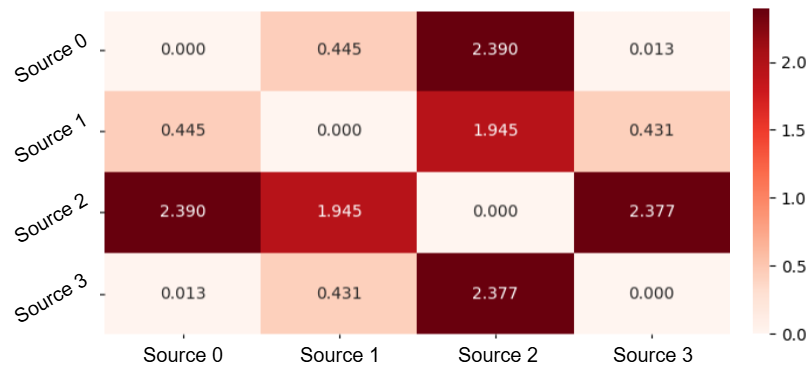}
\caption{Inter-source feature distance matrix.}
\label{fig:matrix}
\vspace{-0.4cm}
\end{figure}

The cross-source analysis reveals varying degrees of domain shift challenges across institutions. Sources 0 and 3 achieve consistently high performance, with minimal validation-to-test degradation, indicating effective domain adaptation. Source 1 shows a moderate performance drop (13.76\%), suggesting more substantial institutional artifacts. Source 2 presents the most challenging scenario due to class imbalance, though performance remains stable between validation and test sets. The outcomes are consistent with~\ref{fig:matrix}; predicting becomes increasingly challenging when the feature distance from other sources is greater.

\subsection{Competition Results}
To validate the practical effectiveness of our preprocessing pipeline in a real-world, multi-institutional setting, we participated in the PHAROS-AFE-AIMI Competition, which serves as an independent benchmark for COVID-19 CT classification; results have shown in Table~\ref{tab:source_analysis} .

For our submission, we employed a simple ensemble strategy. We averaged the prediction probabilities from two models—an EfficientNet-B3~\cite{efficientnet} and a Swin Transformer~\cite{swin}—both of which were independently trained on data processed with our complete SSFL++ and KDS pipeline. This ensemble approach achieved \textbf{first place} in the competition, with a macro-averaged F1-score of 77.6\% on the hidden test set, demonstrating the robustness and efficacy of our method.
\section{Conclusion}
\label{sec:conclusion}

This study demonstrates the effectiveness of a preventive, input-space standardization pipeline for improving domain robustness in multi-source CT classification. Our approach, which combines Spatial-Slice Feature Learning (SSFL++) and Kernel-Density-based Slice Sampling (KDS), performs spatial and temporal standardization to reduce inter-source variance. This process maps disparate inputs into a consistent target space, which mitigates domain shift at its source and simplifies the network's learning task. Our analysis illuminated the fundamental trade-off between local discriminability and cross-source generalization, confirming that both standardization components are necessary for optimal performance. Furthermore, its practical utility was demonstrated by achieving first place in the PHAROS-AFE-AIMI Competition. The results affirm that addressing domain shift at the input level is a powerful and effective strategy for multi-source medical imaging.

{
    \small
    \bibliographystyle{ieeenat_fullname}
    \bibliography{main}
}

\end{document}